# Statistics of microcavity polaritons under non-resonant excitation

Paolo Schwendimann and Antonio Quattropani
Institute of Theoretical Physics. Ecole Polytechnique Fédérale de Lausanne.
CH 1015 Lausanne-EPFL, Switzerland

A model describing amplification and quantum statistics of the exciton polaritons with $\mathbf{k}=0$ in a non-resonantly excited semiconductor quantum well embedded in a microcavity is presented. Exploiting the bottleneck effect for exciton polaritons, it is assumed that the polaritons with $\mathbf{k} \neq 0$ act as a reservoir. The time evolution of the $\mathbf{k}=0$ polaritons is described by a master equation, in which two- and one-polariton transitions between the mode with $\mathbf{k}=0$ and the reservoir are accounted for. The $\mathbf{k}=0$ mode exhibits a threshold depending on the material parameters and on the injected exciton density. Below threshold the quantum statistics of the emission is characteristic of an incoherent process, while above threshold it approaches that of a laser.

71.36.+c, 73.21.Fg

## I INTRODUCTION

In the last few years, non-linear scattering between exciton polaritons has been observed in semiconductor quantum wells embedded in microcavities under resonant as well as non-resonant excitation [1,2]. Resonant non-linear effects have been investigated in experiment [3-5] and in theory [6-8]. It has been shown that for resonant excitation the experiments may be understood in terms of parametric processes (parametric amplification and oscillation) both in a pump and probe configuration and in a configuration in which only the resonant pump is present. It has been also shown that a spontaneous transition from an incoherent emission regime to a coherent one occurs under resonant excitation [9,10]

The emission under non-resonant excitation has also been investigated in experiments. As early as 1998 it has been shown that an amplification regime is present in a CdTe quantum well when the non-resonant pump energy is sufficiently large [11]. The amplified emission is observed in correspondence to the polariton mode with $\mathbf{k}=0$. It has also been shown [12,13] that the amplification is obtained in the strong coupling regime. In more recent years other characteristics of this amplification effect have been investigated experimentally. In particular, the onset of a transition from incoherent to coherent emission has been demonstrated [14] and the quantum statistics of the emitted radiation has been measured [15,16]. This transition has analogies with the transition characterizing the onset of laser action. Furthermore, some exciting results on the spatial coherence of the emission have been recently published [17]. However, in spite of the large number of existing experimental results, the theoretical investigation of the non-resonant amplified emission is still in progress.

The theoretical description of the amplification process has to reproduce the threshold effect observed in the experiments and has to allow calculating the photon statistics of the emitted radiation i.e. the quantum polariton statistics. It should include both the exciton-phonon interaction that is responsible for the energy losses of the excitons generated by non-resonant pumping as well as the exciton-photon and the exciton-exciton interaction leading to the formation of a macroscopically populated polariton state. Most approaches that describe the dynamics of such a complex systems rely on the Boltzmann equation formalism [18-21]. A simplification in the dynamics of the polariton system is obtained by exploiting the existence of a bottleneck effect [22] in the non-resonantly excited polariton system [18, 20]. In fact, for continuous pumping the state of the bottleneck polaritons is assumed to be stationary with the pump thus acting as a reservoir. These Boltzmann equation approaches give important results concerning the evolution of the polariton densities and in particular show the presence of an amplification threshold. However, they don't allow calculating correlation effects and thus the statistics of the emission [15, 17]. Some results on the statistics of polaritons that are obtained in the framework of a master equation approach, have been presented in some recent publications [23, 24]. However, in [23] the polariton-polariton interaction has not been completely included, as we shall indicate in the next Section, while the approach of [24] is based on the polariton-phonon interaction only. A description of the emission statistics based on the polariton-polariton interaction and including the effect of the energy non-conserving terms in the polariton Hamiltonian is still missing.

In this paper we focus on the problem of calculating the photon statistics of the polariton non-linear emission due to the full interaction between polaritons with $\mathbf{k}=0$ and bottleneck polariton with $\mathbf{k}\neq 0$. This interaction includes one- and two-polariton transitions from the bottleneck to the polariton mode with $\mathbf{k}=0$ and vice versa. The polariton-phonon scattering is also active but its contribution to the transitions to the polariton mode with $\mathbf{k}=0$ may be neglected, as we shall discuss in Section 2. The two-polariton transitions play a central role in this approach. They are responsible for the saturation effect that ensure the existence of a stationary solution for the probability distribution and have a direct influence on the values of its moments. These terms have never been considered up to now in the model calculations.

Within a quantum statistical approach we derive a master equation from which the probability of having n polaritons in the mode $\mathbf{k}=0$ is calculated. We present an explicit expression for the amplification threshold allowing its quantitative determination in function of the number of excitons produced by the pump. We show that the emission spectrum experiences a line narrowing at threshold. Furthermore, we analyze the quantum statistics of the emission around the threshold that shows the characteristics of a transition from an incoherent to a coherent emission. Finally we show that above threshold the absolute value of the polariton amplitude becomes different from zero as it happens in a laser system. In this way we are able to characterize completely the polariton emission within the limits of validity of our approximation.

The paper is organized as follows. In Section 2 we present the model and derive the master equation for the evolution of the polaritons with $\mathbf{k}=0$. In Section 3 we discuss the threshold properties of the system and in Section 4 we discuss the statistics of the emission. A discussion of our results is given in Section 5.

**II. THE MODEL**

As we have already anticipated in the Introduction, we consider a situation in which a quantum well embedded in a microcavity is excited by a continuous laser pump near the edge of the conduction band. As it is well known [25] excitons are generated through this excitation process. The dynamics of the excitons is determined by the interaction with the phonons inside the quantum well and by the exciton-exciton interaction. These interactions induce a transfer of the exciton population from the higher energy region in which they are generated into lower energy states. During this process, the excitons accumulate in the flatter region of their energy dispersion outside of the optical region. This region is characterized by values of the wave vector $k$ of the order of $k_0 = E_{exc}(0)\sqrt{\varepsilon_\infty}/\hbar c$, where $\varepsilon_\infty$ is the dielectric constant of the material and $E_{exc}(0)$ is the exciton energy at $\mathbf{k}=0$. Since the transitions into modes with wave vectors $k < k_0$ are characterized by a small density of states, the accumulation of excitons in the energy region around $k_0$ gives rise to a bottleneck effect [22]. This behavior of the excitons is schematically presented in Fig. 1. We are interested in the transitions from the bottleneck region into the polariton mode with $\mathbf{k}=0$. Therefore, in the following we shall not discuss the dynamics of the excitons having energies above that of the bottleneck region.

We describe the system in the polariton operator representation [8]. Two scattering processes may contribute to the building up of a population of polaritons with $\mathbf{k}=0$: polariton-polariton scattering and polariton-phonon scattering. As we shall discuss at the end of this Section, the main process contributing to the transitions to the polariton mode with $\mathbf{k}=0$ is exciton-exciton scattering. Therefore, in the following the exciton-phonon interaction shall not be considered.

We start with the Hamiltonian describing the scattering between pairs of exciton-polaritons in which momentum is conserved. It reads [6]

$$H = \sum_\mathbf{k} \hbar\omega_\mathbf{k} P_\mathbf{k}^+ P_\mathbf{k} + \frac{1}{2}\sum_{\mathbf{k},\mathbf{k'},\mathbf{q}} V_{\mathbf{k},\mathbf{k'},\mathbf{q}} P_{\mathbf{k}+\mathbf{q}}^+ P_{\mathbf{k'}-\mathbf{q}}^+ P_\mathbf{k} P_{\mathbf{k'}} \tag{2.1}$$

with

$$V_{\mathbf{k},\mathbf{k'},\mathbf{q}} = \frac{6e^2\lambda_X}{A\varepsilon} X_\mathbf{k} X_{\mathbf{k'}} X_{\mathbf{k}+\mathbf{q}} X_{\mathbf{k'}-\mathbf{q}} + \frac{\hbar\Omega_R 16\pi\lambda_X^2}{7A} X_\mathbf{k} X_{\mathbf{k'}-\mathbf{q}} \left( X_{\mathbf{k'}} |C_{\mathbf{k}+\mathbf{q}}| X_{\mathbf{k}+\mathbf{q}} + |C_{\mathbf{k'}}| X_{\mathbf{k}+\mathbf{q}} \right).$$

Here $A$ is the quantization area, $\lambda_x$ is the exciton radius and $\Omega_R$ is the Rabi frequency. The coefficients $X_\mathbf{k}$ and $C_\mathbf{k}$ are the Hopfield coefficients for the exciton and the photon components of the polariton respectively. The polariton operators $P_\mathbf{k}, P_\mathbf{k}^+$ may be assumed to obey Bose commutation relations because we are considering an excitation regime in which the density of the excitons is smaller than their saturation density $n_{sat}$ [26]. We remark that the pumping process is not included in the Hamiltonian (2.1) and we shall discuss this point after having specified our model in more detail.

We are interested in the scattering processes leading to transitions from the bottleneck region to the polariton state with $\mathbf{k}=0$. Therefore, we rewrite (2.1) by separating out of it the terms containing the polariton operators $P_0, P_0^+$. We thus obtain

$$H = H_0 + \sum_{\mathbf{k}\neq 0,} W_{\mathbf{k},-\mathbf{k}}(P_0^+ P_0^+ P_\mathbf{k} P_{-\mathbf{k}} + h.c.) + \sum_{\mathbf{k},\mathbf{k'},} W_{\mathbf{k},\mathbf{k'}}(P_0^+ P_{\mathbf{k}+\mathbf{k'}}^+ P_\mathbf{k} P_{\mathbf{k'}} + P_\mathbf{k}^+ P_{\mathbf{k'}}^+ P_{\mathbf{k}+\mathbf{k'}} P_0) +$$
$$W_0 P_0^+ P_0^+ P_0 P_0 + \sum_{\mathbf{k}\neq 0,} W_\mathbf{k}, P_\mathbf{k}^+ P_\mathbf{k} P_0^+ P_0 + \text{terms not containing } P_0 \tag{2.2a}$$

with

$$H_0 = \hbar\omega_0 P_0^+ P_0 + \sum_\mathbf{k} \hbar\omega_\mathbf{k} P_\mathbf{k}^+ P_\mathbf{k} . \tag{2.2b}$$

The coefficients that appear in (2.2) are defined as

$$W_{\mathbf{k},} = \frac{1}{2}(V_{\mathbf{k},0,-\mathbf{k}} + V_{\mathbf{k},0,\mathbf{k}}) \ , \qquad (2.3a)$$

$$W_{\mathbf{k},-\mathbf{k}} = \frac{1}{2}(V_{0,0,\mathbf{k}} + V_{0,0,-\mathbf{k}}) \ , \qquad (2.3b)$$

$$W_{\mathbf{k},\mathbf{k}'} = \frac{1}{2}(V_{0,\mathbf{k}+\mathbf{k}',\mathbf{k}} + V_{\mathbf{k}+\mathbf{k}',0,-\mathbf{k}}) \ , \qquad (2.3c)$$

$$W_0 = \frac{1}{2}V_{0,0,0} \ . \qquad (2.3d)$$

We don't write explicitly the terms not containing $P_0$, $P_0^+$ in (2.2) because we shall not need them in the following. The Hamiltonian (2.2) contains several interaction terms that we now discuss in some detail. The interaction term containing two polaritons with $\mathbf{k}=0$ describes processes in which two polaritons with $\mathbf{k}=0$ are created (annihilated) and a pair of polaritons with opposite wave vectors $\mathbf{k} \neq 0$ is annihilated (created). These terms correspond to processes that conserve momentum but not energy. They shall play an important role in the following and their contributions have not been considered in the existing literature on the subject [18-20, 22-24] in particular they are missing in [23]. The terms containing only one polariton operator with $\mathbf{k}=0$ describe the creation (annihilation) of one polariton with $\mathbf{k}=0$ through the scattering between polaritons with different wave vectors. Finally, the Hamiltonian (2.1) contains a self-interaction term between the polaritons at $\mathbf{k}=0$ and a term that renormalizes the polariton energy due to the interaction with all polaritons with a wave vector different from zero.

The presence of the bottleneck effect together with the choice of a continuous pump excitation allows us to assume that the polaritons in the bottleneck region are in a stationary state with the pump. Due to the small density of states, we also assume that the transition from the bottleneck into the polariton state with $\mathbf{k}=0$ is a direct transition. These assumptions allow us to think of the bottleneck polaritons as a reservoir, which couples to the polaritons with $\mathbf{k}=0$ through the processes described by (2.2).

In order to describe the dynamics of the polaritons with $\mathbf{k}=0$ and their statistics, we make use of an approach based on the density operator description of a multicomponent system. Starting point for this description is the Liouville-von Neumann equation

$$i\hbar \frac{d\rho}{dt} = [H,\rho] + iD\rho \equiv L\rho \ . \qquad (2.4)$$

Here $D$ is a dissipation operator describing the microcavity losses and having the following properties

$$D\rho = \sum_{\mathbf{k}} \gamma_{\mathbf{k}} \left( [P_{\mathbf{k}}\rho, P_{\mathbf{k}}^+] + [P_{\mathbf{k}}, \rho P_{\mathbf{k}}^+] \right) , \qquad (2.5a)$$

$$DP_{\mathbf{k}} = -\gamma_{\mathbf{k}} P_{\mathbf{k}} \ . \qquad (2.5b)$$

In order to obtain from (2.4) an equation involving the reduced density operator $\rho_0 = Tr P \rho$ describing the polaritons with $\mathbf{k}=0$ only, we use a projector formalism [27]. We define

$$P\rho = \rho^{stationary}Tr_{res}\rho \;, \tag{2.6a}$$

$$P^2 = PP(1-P) = (1-P)P = 0 \;, \tag{2.6b}$$

$$(1-P)\rho(0) = 0 \;. \tag{2.6c}$$

We notice that $\rho^{stationary}$ is the equilibrium density operator for the polariton modes with $\mathbf{k} \neq 0$ and we assume $\rho(0) = \rho_0(0) \otimes \rho^{stationary}$ as initial condition. The trace appearing in (2.6) is understood being performed over all polariton states with $\mathbf{k} \neq 0$. The derivation of a master equation for $\rho_0(t)$ is carried out using standard techniques. The main approximation consists in retaining only terms that are of second order in the interaction between the modes with $\mathbf{k} = 0$ and those with $\mathbf{k} \neq 0$. This approximation is consistent with the assumption that the polaritons with $\mathbf{k} \neq 0$ form a reservoir. Retaining higher order terms in the interaction would imply that, contrary to our assumption, the state of the reservoir is changed due to the interaction with the polaritons with $\mathbf{k} = 0$. We obtain

$$i\hbar \frac{dP\rho}{dt} = PLP\rho - \frac{i}{\hbar}\int_0^t PLL(\tau)P\rho(t-\tau)\,d\tau \;, \tag{2.7a}$$

$$L(\tau)P\rho(t-\tau) = \exp\bigl[-i/\hbar(L_0 + iD)\tau\bigr]LP\rho(t-\tau) \;, \tag{2.7b}$$

where $L_0\rho = [H_0, \rho]$. The next step consists in explicitly evaluating the traces and in performing a Markov approximation in the time integral that appears in (2.7). This last approximation is justified because, due to the presence of an external continuous pump field, the evolution of the polaritons with $\mathbf{k} = 0$ will be slower than all relaxation effects in the reservoir. We present some details of this derivation in the Appendix. The resulting master equation for the reduced density operator $\rho_0$ reads

$$\begin{aligned}
\hbar \frac{d}{dt}\rho_0(t) = &-i\bigl[W_0 P_0^{+2} P_0^2, \rho_0(t)\bigr] - i\bigl[\hbar\hat{\omega}_0 P_0^+ P_0, \rho_0(t)\bigr] + \\
&\Gamma_1 \bigl([P_0^2 \rho_0(t), P_0^{2+}] + [P_0^2, \rho_0(t)P_0^{2+}]\bigr) + \Delta_1 \bigl([P_0^{+2}\rho_0(t), P_0^2] + [P_0^{+2}, \rho_0(t)P_0^2]\bigr) + \\
&\Gamma_2 \bigl([P_0 \rho_0(t), P_0^+] + [P_0, \rho_0(t)P_0^+]\bigr) + \Delta_2 \bigl([P_0^+ \rho_0(t), P_0] + [P_0^+, \rho_0(t)P_0]\bigr) + \\
&\gamma_0 \bigl([P_0 \rho_0(t), P_0^+] + [P_0, \rho_0(t)P_0^+]\bigr) \;.
\end{aligned} \tag{2.8}$$

The coefficients in (2.8) are expressed in terms of the expectation values of the polariton numbers with $\mathbf{k} \neq 0$ as

$$\Gamma_1 = \frac{1}{\hbar} \sum_{\mathbf{k} \neq 0,} \operatorname{Re} G^{(+)}_{\mathbf{k},-\mathbf{k}} W_{\mathbf{k},-\mathbf{k}} W_{\mathbf{k},-\mathbf{k}} \left( 2 \langle P^+_\mathbf{k} P_\mathbf{k} \rangle \langle P^+_{-\mathbf{k}} P_{-\mathbf{k}} \rangle + \langle P^+_{-\mathbf{k}} P_{-\mathbf{k}} \rangle + \langle P^+_\mathbf{k} P_\mathbf{k} \rangle \right), \qquad (2.9a)$$

$$\Delta_1 = \frac{2}{\hbar} \sum_{\mathbf{k} \neq 0,} \operatorname{Re} G^{(+)}_{\mathbf{k},-\mathbf{k}} W_{\mathbf{k},-\mathbf{k}} W_{\mathbf{k},-\mathbf{k}} \langle P^+_\mathbf{k} P_\mathbf{k} \rangle \langle P^+_{-\mathbf{k}} P_{-\mathbf{k}} \rangle, \qquad (2.9b)$$

$$\Gamma_2 = \frac{1}{\hbar} \sum_{\mathbf{k},\mathbf{k}',} \operatorname{Re} G^{(+)}_{\mathbf{k},\mathbf{k}',\mathbf{k}+\mathbf{k}'} W_{\mathbf{k},\mathbf{k}'} \left( W_{\mathbf{k},\mathbf{k}'} + W_{\mathbf{k}',\mathbf{k}} \right) \times$$
$$\left[ \langle P^+_{\mathbf{k}+\mathbf{k}'} P_{\mathbf{k}+\mathbf{k}'} \rangle \left( \langle P^+_\mathbf{k} P_\mathbf{k} \rangle + \langle P^+_{\mathbf{k}'} P_{\mathbf{k}'} \rangle + 1 + \langle P^+_\mathbf{k} P_\mathbf{k} \rangle \langle P^+_{\mathbf{k}'} P_{\mathbf{k}'} \rangle \right) \right], \qquad (2.10a)$$

$$\Delta_2 = \frac{1}{\hbar} \sum_{\mathbf{k},\mathbf{k}',} \operatorname{Re} G^{(+)}_{\mathbf{k},\mathbf{k}',\mathbf{k}+\mathbf{k}'} W_{\mathbf{k},\mathbf{k}'} \left( W_{\mathbf{k},\mathbf{k}'} + W_{\mathbf{k}',\mathbf{k}} \right) \times$$
$$\left[ \langle P^+_{\mathbf{k}+\mathbf{k}'} P_{\mathbf{k}+\mathbf{k}'} \rangle \langle P^+_\mathbf{k} P_\mathbf{k} \rangle \langle P^+_{\mathbf{k}'} P_{\mathbf{k}'} \rangle + \langle P^+_\mathbf{k} P_\mathbf{k} \rangle \langle P^+_{\mathbf{k}'} P_{\mathbf{k}'} \rangle \right]. \qquad (2.10b)$$

The quantities $G^{(+)}_{ij}$ that appear in the definitions are explicitly given in the Appendix. The frequency $\hat{\omega}_0$ is the frequency of the polaritons with $\mathbf{k} = 0$ renormalized by the interaction with the reservoir. The coefficients $\Gamma_1$ and $\Delta_1$ are the rates for processes in which a polariton pair is subtracted from or injected into the mode with $\mathbf{k} = 0$, respectively due to the exchange of two polaritons with opposite wave vector with the reservoir. The coefficients $\Gamma_2$ and $\Delta_2$ are the rates for subtraction and injection of one polariton into the mode $\mathbf{k} = 0$ respectively due to the interaction with the reservoir. The characteristics of these coefficients are discussed in the next Section.

At the beginning of this Section we stated that the contributions of the polariton-phonon scattering to the amplification process might be neglected. It may be shown in the framework of our master equation approach, that phonon-exciton scattering gives a contribution to the rates $\Gamma_2$ and $\Delta_2$. This contribution is shown to be small, its only influence on the dynamics consisting in slightly shifting the threshold for amplification towards higher values [28]. Otherwise it doesn't lead to any important deviation from the behavior discussed in the following Sections.

### III. THE THRESHOLD CONDITION

The solution of the master equation (2.8) will be presented and discussed in the next Section. In order to get some preliminary insight into the processes described by (2.8) we consider the equation for the expectation value of the polariton number $\langle P^+_0 P_0 \rangle$, which is easily derived from (2.8) and reads

$$\hbar \frac{d}{dt} \langle P^+_0 P_0 \rangle = (8\Delta_1 + 2\Delta_2 - 2\Gamma_2 - 2\gamma_0) \langle P^+_0 P_0 \rangle -$$
$$4(\Gamma_2 - \Delta_1) \langle P^{+2}_0 P^2_0 \rangle + (8\Delta_1 + 2\Delta_2). \qquad (3.1)$$

The equation (3.1) is the first equation in a hierarchy and thus cannot be solved as it is. However, we can learn some interesting features of the polariton emission process from it. The coefficient of the term $\langle P^+_0 P_0 \rangle$ in (3.1) consists of the difference between the polariton injection terms $\Delta_1$ and $\Delta_2$ and the polariton losses characterized by the rates $\Gamma_2$ and $\gamma_0$. As we shall show in the following, this difference changes its sign as a function of the pumping and may describe an amplification process when the threshold condition

$$(4\Delta_1 + \Delta_2 - \Gamma_2 - \gamma_0) > 0 \tag{3.2}$$

is satisfied. The inequality (3.2) is reminiscent of the relation between gain and losses in a laser around threshold. Indeed, the equation (3.1) is formally analogous to the equation that describes the evolution of the photon number of a continuously pumped one-mode laser around threshold [29]. It is important to remark that this analogy is only formal. In fact, the next equation in the hierarchy i.e., the evolution of the correlation $\langle P_0^{+2} P_0^2 \rangle$ reads

$$\hbar \frac{d}{dt} \langle P_0^{+2} P_0^2 \rangle = (24\Delta_1 + 4\Delta_2 - 4(\Gamma_1 + \Gamma_2) - 4\gamma_0) \langle P_0^{+2} P_0^2 \rangle - $$
$$8(\Gamma_1 - \Delta_1) \langle P_0^{+3} P_0^3 \rangle + (64\Delta_1 + 8\Delta_2) \langle P_0^+ P_0 \rangle + 8\Delta_1 . \tag{3.3}$$

We notice that the source term $8\Delta_1$ in (3.3) is absent in the equivalent equation for the one mode laser. It originates in the two-polariton processes and shows their importance in determining the quantum statistics of the polariton emission. In order to discuss the condition (3.2), we need explicit expressions for the expectation values $\langle P_\mathbf{k}^+ P_\mathbf{k} \rangle$ that appear in the definitions (2.9) and (2.10) of $\Gamma_i$ and $\Delta_i$ (i = 1,2).

In Section 2 we have assumed that the polaritons in the bottleneck region act as a reservoir whose stationary state is determined by the continuous pump. This stationary state arises as a consequence of the balance between the energy of the exciton injected into the system through the pump and the exciton energy dissipation via phonon-exciton scattering. Instead of giving a detailed picture of this process, we suppose that for a given pump value the density of the excitons in the bottleneck is a constant. We then simulate the stationary polariton state by a thermal equilibrium state characterized by a temperature of the order of the lattice temperature and by a chemical potential µ. The average polariton number for a given wave vector **k** reads

$$\langle P_\mathbf{k}^+ P_\mathbf{k} \rangle = \frac{1}{\exp[\beta(E_{pol}(\mathbf{k}) - \mu)] - 1} . \tag{3.4}$$

In order to obtain an explicit expression for the chemical potential, we exploit the assumption that the polariton density in the bottleneck is a constant. The chemical potential is then calculated through the relation

$$N = \sum_\mathbf{k} \frac{1}{\exp[\beta(E_{pol}(\mathbf{k}) - \mu)] - 1} . \tag{3.5}$$

Here $N$ is the particle number. Since we are considering a situation in which most polaritons are accumulated in the bottleneck region, we assume that the energy dispersion of the bottleneck polaritons may approximated by the one of the excitons i.e. by

$$E_{exc}(\mathbf{k}) = E_{exc}(0) + \frac{\hbar^2 \mathbf{k}^2}{2M} . \tag{3.6}$$

Here $M$ is the total mass of the exciton. The sum in (3.5) is now transformed in an integral and we get for the chemical potential the expression

$$\mu = E_{exc}(0) + \frac{1}{\beta}\log\left(1 - \exp\left[-4\pi\beta\hbar^2 N / 2MA\right]\right). \tag{3.7}$$

Having specified the expectation values of the bottleneck polariton numbers, we calculate the explicit expressions for the rates that appear in (2.8) by transforming the sums in (2.9) and (2.10) into integrals. We notice that the rates $\Gamma_2$ and $\Delta_2$ are not independent quantities, but are related by the general thermodynamic relation

$$\Delta_2 = \Gamma_2 \exp\left[-\beta(E_{pol}(0) - \mu)\right], \tag{3.8}$$

as it can be easily verified from the definitions (2.9) and (2.10). The relation (3.8) shows that depending on the sign of the difference $E_{pol}(0) - \mu$, the injection rate $\Delta_2$ may become larger than the dissipation rate $\Gamma_2$ when the condition

$$\mu > E_{pol}(0) \tag{3.9}$$

is satisfied. We now rewrite the equation (3.2) making use of (3.8) as

$$\left(4\Delta_1 + \Gamma_2 \left(\exp\left[-\beta(E_{pol}(0) - \mu)\right] - 1\right) - \gamma_0\right) > 0. \tag{3.10}$$

Equation (3.10) gives the threshold condition for the polariton amplification as a function of the polariton density in the bottleneck i.e. of the external pump. In Fig. 2 we show the dependence of (3.10) on the external pump for a CdTe quantum well embedded in a microcavity. We use the following material parameters for a CdTe quantum well: $E_{pol}(0) = E_{exc}(0) - \hbar\Omega_{Rab}$, with $E_{exc}(0) = 1680\,meV$ and $2\hbar\Omega_{Rab} = 7\,meV$, $\varepsilon_\infty = 7.4$, the total mass of the exciton $M = 0.296 m_e$, the quantization area $A = 10^{-6}\,cm^2$, the polariton-polariton interaction constant $W = 3.6\,10^{-5}\,meV$, and the temperature $T = 12\,K$. The gain curve in Fig. 2 allows distinguishing between three different regimes. When the condition (3.9) is not satisfied, no gain is present in the system. When the condition (3.9) is satisfied, the system shows a positive internal gain, which however doesn't balance the losses of the microcavity. In this case the system doesn't show amplification but the spontaneous emission line will exhibit a narrowing effect. Finally, when the condition (3.10) is satisfied, amplification will occur. We notice that the line narrowing $\left(d = 2\,10^9\,cm^{-2}\right)$ and the amplification $\left(d = 2.27\,10^{10}\,cm^{-2}\right)$ appear for values of the density smaller than the saturation density $d = 6.7\,10^{11}\,cm^{-2}$ thus justifying the bosonic description of excitons that underlies our approach. The density, at which the threshold occurs, is of the same order of magnitude of the one of $d = 6\,10^{10}\,cm^{-2}$ found in the experiments [14]. We notice that, the threshold values depend on the chosen value for the cavity losses i.e. $\gamma_0 = 0.1\,meV$. By using material parameters characteristic of GaAs a similar behavior of the gain is found. The only difference with the CdTe calculation concerns the position of the threshold that is found for a density $d = 1.36\,10^{10}\,cm^{-2}$ with the same value of $\gamma_0$ and a saturation density $n_{sat} = 10^{11}\,cm^{-2}$.

## IV. STATISTICAL PROPERTIES OF THE POLARITON EMISSION

In order to discuss the statistics of the polariton emission, we have to solve the master equation (2.8). To this end, we rewrite it in terms of its matrix elements taken on the Fock states of the polaritons. The diagonal matrix elements of the reduced density operator obey the equation

$$\hbar \frac{d}{dt}\rho_n(t) = 2(n+1)(n+2)\Gamma_1\rho_{n+2}(t) + 2(n+1)(\Gamma_2 + \gamma_0)\rho_{n+1}(t) -$$
$$\left[2n(\gamma_0 + \Gamma_2) + 2n(n-1)\Gamma_1 + 2(n+1)(n+2)\Delta_1 + 2(n+1)\Delta_2\right]\rho_n(t) +$$
$$2n\Delta_2\rho_{n-1}(t) + 2n(n-1)\Delta_1\rho_{n-2}(t) \ . \tag{4.1}$$

In order to simplify the notation we have introduced $\rho_n = \langle n|\rho_0|n\rangle$ in (4.1). The off-diagonal matrix elements of the reduced density operator satisfy the equation

$$\hbar \frac{d}{dt}\rho_{n,m}(t) =$$
$$\Gamma_1\left[2\sqrt{(n+1)(n+2)(m+1)(m+2)}\rho_{n+2,m+2}(t) - \big(n(n+1) + m(m+1)\big)\rho_{n,m}(t)\right] +$$
$$\Delta_1\left[2\sqrt{n(n-1)m(m-1)}\rho_{n-2,m-2}(t) - \big((n+2)(n+1) + (m+2)(m+1)\big)\rho_{n,m}(t)\right] +$$
$$(\Gamma_2 + \gamma_0)\left[2\sqrt{(n+1)(m+1)}\rho_{n+1,m+1}(t) - (n+m)\rho_{n,m}(t)\right] +$$
$$\Delta_2\left[2\sqrt{nm}\rho_{n-1,m-1}(t) - (n+m+2)\rho_{n,m}(t)\right] -$$
$$i\left[W_0\big(n(n-1) - m(m-1)\big) + \hbar\hat{\omega}_0(n-m)\right]\rho_{n,m}(t) \ . \tag{4.2}$$

We notice that the off diagonal elements of the reduced density matrix evolve in time independently of the diagonal ones. Since for t=0 all the off-diagonal elements are zero, they will be zero all the time. On the contrary, the diagonal elements evolve with the initial conditions $\rho_{n=0} = 1$ and $\rho_{n\neq 0} = 0$ ensuring that a solution different from zero exists in this case. Solving (4.1) for a finite value of *n*, and with the condition $Tr\rho = 1$, shows that for large values of time, a stationary regime is obtained. The existence of a stable stationary state for $\rho_n$ below as well as above threshold is related to the presence of the two-polariton terms in (4.1). Without these terms equation (4.1) reduces to the master equation for a harmonic oscillator whose solution diverges in time when $\Delta_2 > (\Gamma_2 + \gamma_0)$. These characteristics allow interpreting the two-polariton terms in (4.1) as being responsible for saturation as it has been anticipated in the previous Section.

Equation (4.2) is useful when calculating the polariton spectrum i.e. the Fourier time transform of the correlation $\langle P_0^+(t)P_0(0)\rangle$. It can be shown [30] that this correlation may be expressed as

$$\langle P_0^+(t)P_0(0)\rangle = \sum_n n\rho_n(0)\rho_{n,n-1}(t) \ . \tag{4.3}$$

The stationary solution $\rho_n^{stat}$ for $\rho_n(t)$ is the solution of the equation

$$2(n+1)(n+2)\Gamma_1\rho^{stat}_{n+2} + 2(n+1)(\Gamma_2+\gamma_0)\rho^{stat}_{n+1} -$$
$$\left[2(n+1)(n+2)\Delta_1 + 2(n+1)\Delta_2 + 2n(\gamma_0+\Gamma_2) + 2n(n-1)\Gamma_1\right]\rho^{stat}_n +$$
$$2n\Delta_2\rho^{stat}_{n-1} + 2n(n-1)\Delta_1\rho^{stat}_{n-2} = 0 \ . \tag{4.4}$$

In equation (4.4) there is a balance between the transitions from the states with $n+2$ polaritons (two-polariton dissipation) and from the states with $n-2$ polaritons (two-polariton injection) into the state with n polaritons and similarly between the transitions from the states with $n+1$ polaritons and with $n-1$ polaritons into the state with n polaritons. We solve (4.4) numerically by a truncation procedure, which takes advantage of the condition $Tr\rho = 1$. In the following we shall discuss the statistics of the polariton emission from the CdTe quantum well embedded in a microcavity introduced in the previous Section. We shall present the polariton distribution as well as its lowest order normalized factorial moments defined as

$$M_p = \langle n(n-1)......(n-p+1)\rangle / \langle n\rangle^p \ . \tag{4.5}$$

The results of this calculation are presented in Figs. 3-6. In Fig. 3 we show how the polariton distribution varies with the density of excitons induced by the external pump. Below threshold the polariton distribution $\rho_n$ is centered on the value $n=0$ for the polariton number and resembles to a geometrical (thermal) distribution, which is known to characterize an incoherent emission process. At threshold the distribution although always centered on $n=0$ has changed its tangent at $n=0$. We notice that at threshold the dominating effect is due to the undamped spontaneous emission as it may be seen from (3.1) and (3.3). Above threshold the distribution changes completely: its maximum value is displaced towards higher values of $n$ and its form resembles a Poissonian distribution. The global behavior of the probability distribution is characteristic of a process in which there is a transition from an incoherent to a coherent emission. This characteristic is evident when considering the normalized factorial moments (4.5) that are presented in Fig. 4 and Fig. 5 and compared with the moments of the photon distribution from a laser whose characteristic parameters (gain, losses and saturation) have the same values as the equivalent quantities in the polariton model. The first moment is presented in Fig. 4 as a function of the exciton density and clearly shows amplification at threshold. Compared with the laser emission it shows a smaller amplification due to the larger noise in our system. The second moment is presented in Fig. 5. As expected from the form of the probability distribution, below threshold the values of second moment are characteristic of an incoherent process ($M_2 \geq 2$). Above threshold the second moment approaches to the value characteristic for coherent emission ($M_2 = 1$). However, coherence is obtained for larger values of the gain than in the laser case. The presence of an amplification threshold is also evident in the behavior of the linewidth of the emission. As it is shown in Fig. 6, the linewidth of the polariton strongly decreases for densities near threshold and attains its minimum at threshold remaining afterward much smaller that it was below threshold.

We expect the expectation value of the polariton amplitude $\langle P_0 \rangle$ to vanish for any time because the off diagonal elements of the reduced density operator, by means of which it is calculated, are always zero. However, the absolute value of $\langle P_0 \rangle$ may be different from zero above threshold. This behavior is suggested when discussing the evolution of $\langle P_0 \rangle$ in the mean field approximation. The time evolution of $\langle P_0 \rangle$ is described by

$$i\hbar \frac{d}{dt}\langle P_0 \rangle = \hbar\hat{\omega}_0 \langle P_0 \rangle + i(4\Delta_1 + \Delta_2 - \Gamma_2 - \gamma_0)\langle P_0 \rangle + 2(W_0 - i(\Gamma_1 - \Delta_1))\langle P_0^+ P_0^2 \rangle \ . \quad (4.6)$$

In the mean field approximation (4.6) becomes

$$i\hbar \frac{d}{dt}\langle P_0 \rangle = \hbar\hat{\omega}_0 \langle P_0 \rangle + i(4\Delta_1 + \Delta_2 - \Gamma_2 - \gamma_0)\langle P_0 \rangle +$$
$$2(W_0 - i(\Gamma_1 - \Delta_1))|\langle P_0 \rangle|^2 \langle P_0 \rangle \ . \quad (4.7)$$

We look for a solution of (4.7) in the form $\langle P_0(t) \rangle = |\langle P_0 \rangle| \exp(-i\omega t)$, with real frequency $\omega$. Such a solution of (4.7) exists provided that $|\langle P_0 \rangle|$ is given by

$$|\langle P_0 \rangle| = 0 \quad (4.8a)$$

or

$$|\langle P_0 \rangle| = \sqrt{\frac{4\Delta_1 - (\Gamma_2 + \gamma_0 - \Delta_2)}{2(\Gamma_1 - \Delta_1)}} \ . \quad (4.8b)$$

The solution (4.8a) is stable below threshold only, while the solution (4.8b) is stable only above threshold. We verify this result by calculating the expectation value of the absolute value of the amplitude using (4.4). To this end, we introduce the operator describing the absolute value of the amplitude [31]

$$|P_0| = \sqrt{P_0^+ P_0 + 1} \ . \quad (4.9)$$

The expectation value of this quantity is presented in Fig. 7 as a function of the exciton density. From Fig. 7 it is clear that the behavior predicted by the mean field result is also found in the exact solution of the stationary equation (4.4). The results presented in Figs. 3-7 clearly indicate that above threshold a macroscopically populated polariton state appears. The polariton distribution pertaining to this state has characteristics similar to the ones of the photons in a one-mode laser, although the mechanism leading to the emission is different

## V. CONCLUSIONS

We now summarize the results presented in the former Sections. We have derived a master equation description of the dynamics of the polaritons with $\mathbf{k} = 0$ that includes both one- and two-polariton transitions between the mode $\mathbf{k} = 0$ and the reservoir (bottleneck). In particular, the two-polariton transitions lead to saturation effects in the dynamics of the polaritons and guarantee the emerging of a stationary solution. We have shown that the amplification threshold is naturally contained in our approach and we have derived its explicit expression from which quantitative calculation of the threshold value in a realistic material may be carried out. Furthermore we have presented the stationary polariton number distribution and the emission spectrum as a function of the exciton density in the system below as well as above threshold. Both show features that are characteristic of a transition from an incoherent to a coherent emission. We have also shown that above threshold the expectation value of the absolute value of the polariton field amplitude is different from zero, a behavior characteristic of the transition from incoherence to coherence. Although the polariton number distribution is similar to the one of the one-mode laser, the behavior of the moments of the polariton distribution is different from the one of the photon distribution in a laser. As we have shown in Section 3, equation (3.4) describing the evolution of the second moment of the polariton distribution differs from that of the laser by a source term related to the two-polariton transitions. Since in our example the numerical value of the source term is very small compared to the ones of the one-polariton transitions, the difference between the polariton and the laser emission is not expected to be dramatic. However, for a different choice of the system parameters a part from the formal differences, a more relevant difference between the polariton emission and that of a laser may appear.

Finally, we briefly discuss the limits of our model. We have implicitly assumed that all polaritons with $\mathbf{k} \neq 0$ pertain to the reservoir. In fact, a more refined model should consider the polariton states lying between the bottleneck and the state with $\mathbf{k} = 0$ to be involved in the dynamics. It may be shown that combining the master equation techniques outlined in Section 2 with the approximation that no correlations between the polaritons with $\mathbf{k} = 0$ and the ones with $\mathbf{k} \neq 0$ and $\mathbf{k} < \mathbf{k}_0$ are considered, a new master equation for $\rho_0$ only is obtained [28]. This new master equation has the same form as (2.8), but for the rates appearing in this new equation. This new rates, denoted by $\hat{\Gamma}_i(t)$ and $\hat{\Delta}_i(t)$ (i=1,2), are different from the ones given by (2.9) and (2.10) because they contain the dynamical expectation values of the number of polaritons with $\mathbf{k} \neq 0$ and $\mathbf{k} < \mathbf{k}_0$. However, the stationary regime is described by (4.4) also in this case, the rates $\Gamma_i$ and $\Delta_i$ (i=1,2) being replaced by the stationary values of $\hat{\Gamma}_i(t)$ and $\hat{\Delta}_i(t)$ (i=1,2). Therefore, we expect that the main results on the statistics in the stationary state presented in Section 4 will also hold when more polariton modes below the bottleneck are included.

**APPENDIX**

In order to avoid complicated expression let us rewrite the Hamiltonian (2.2) as

$$H = H_{01} + H_{02} + H_3 + H_1 + H_2 + H_4, \quad (A1a)$$

$$H_{01} + H_{02} + H_3 + H_4 = \hbar\omega_0 P_0^+ P_0 + \sum_{\mathbf{k}} \hbar\omega_{\mathbf{k}} P_{\mathbf{k}}^+ P_{\mathbf{k}} + \sum_{\mathbf{k}\neq 0,} W_{\mathbf{k}} P_{\mathbf{k}}^+ P_{\mathbf{k}} P_0^+ P_0 + \frac{1}{2} V_{0,0,0} P_0^+ P_0^+ P_0 P_0 \quad , \quad (A1b)$$

$$H_1 = \sum_{\mathbf{k}\neq 0,} W_{\mathbf{k},-\mathbf{k}} (P_0^+ P_0^+ P_{\mathbf{k}} P_{-\mathbf{k}} + h.c.) \quad , \quad (A1c)$$

$$H_2 = \sum_{\mathbf{k},\mathbf{k}'} W_{\mathbf{k},\mathbf{k}'} (P_0^+ P_{\mathbf{k}+\mathbf{k}'}^+ P_{\mathbf{k}} P_{\mathbf{k}'} + P_{\mathbf{k}}^+ P_{\mathbf{k}'}^+ P_{\mathbf{k}+\mathbf{k}'} P_0) \quad . \quad (A1d)$$

Equation (2.7) is rewritten in explicit form as

$$i\hbar\frac{d}{dt}\rho_0(t) = [(H_{01} + H_3), \rho_0(t)] + [Z P_0^+ P_0, \rho_0(t)] -$$

$$-\frac{i}{\hbar}\int_0^t d\tau Tr_{res}([(H_1 + H_2),[(H_1(\tau) + H_2(\tau)), \rho_{res}\rho_0(\tau, t-\tau)]]) + D\rho_0(t) \quad , \quad (A2a)$$

$$Z = \sum_{\mathbf{k}\neq 0} W_{\mathbf{k}} \langle P_{\mathbf{k}}^+ P_{\mathbf{k}} \rangle \quad . \quad (A2b)$$

The double commutator in (A2) separates into the two double commutators for the two Hamiltonian contributions. There are no contributions of a double commutator containing Hamiltonians with different indexes. This is due to the fact, that in thermal equilibrium the expectation values of the products of polariton operators factorize into products of expectation values of number operators.

The two commutators involving $H_1$ and $H_2$ have the same form namely

$$E(t,\tau) = Tr_{res}\left[(A^+B + B^+A),\left[(A^+(\tau)B(\tau) + B^+(\tau)A(\tau)), \rho_{eq}\rho_0(\tau, t-\tau)\right]\right] =$$
$$-\langle[B,B^+]\rangle(\tau)\left([A\rho_0(\tau,t-\tau),A^+] + [A,\rho_0(\tau,t-\tau)A^+]\right)+$$
$$\langle B^+B\rangle(\tau)\left([A\rho_0(\tau,t-\tau),A^+] + [A^+\rho_0(\tau,t-\tau),A]\right)+$$
$$\left([A,\rho_0(\tau,t-\tau)A^+] + [A,^+\rho_0(\tau,t-\tau)A]\right), \qquad (A3a)$$

$$G_{\mathbf{k},\pm}(\tau) = \exp(\pm i(\omega_0 - \omega_\mathbf{k})\tau - (\gamma_0 + \gamma_\mathbf{k})\tau) \ . \qquad (A3b)$$

where the operators $A$ and $B$ for $H_1$ are $A = P_0^2$ and $B(\tau) = \sum_{\mathbf{k}\neq 0,} W_{\mathbf{k},-\mathbf{k}}(P_\mathbf{k} P_{-\mathbf{k}})G_{\mathbf{k},+}(\tau)G_{-\mathbf{k},+}(\tau)$,

and the operators $A$ and $B$ for $H_2$ are $A = P_0$ and

$$B(\tau) = \sum_{\mathbf{k},\mathbf{k}',} W_{\mathbf{k},\mathbf{k}'} P^+_{\mathbf{k}+\mathbf{k}'} P_\mathbf{k} P_{\mathbf{k}'} G_{\mathbf{k}+\mathbf{k}',-}(\tau) G_{\mathbf{k},+}(\tau) G_{\mathbf{k}',+}(\tau).$$

At this point we perform a Markov approximation: we extend to infinity the upper integration limit in (A2) and set $\rho_0(\tau, t-\tau) \approx \rho_0(t)$. This approximation is justified by assuming that the reduced density operator varies slowly on the time scale characterized by the relaxation times of the polariton modes k. The quantities $G_\pm(t)$ are integrated over time in (A3) and become

$$G^{(-)}_{\mathbf{k},-\mathbf{k}} = \frac{-i(\omega_0 - \omega_\mathbf{k} - \omega_{-\mathbf{k}}) + (\gamma_0 + \gamma_\mathbf{k} + \gamma_{-\mathbf{k}'})/\hbar}{(\omega_0 - \omega_\mathbf{k} - \omega_{-\mathbf{k}})^2 + (\gamma_0 + \gamma_\mathbf{k} + \gamma_{-\mathbf{k}})^2/\hbar^2} \ ,$$

$$G^{(+)}_{\mathbf{k},-\mathbf{k}} = \frac{i(\omega_0 - \omega_\mathbf{k} - \omega_{-\mathbf{k}}) + (\gamma_0 + \gamma_\mathbf{k} + \gamma_{-\mathbf{k}'})/\hbar}{(\omega_0 - \omega_\mathbf{k} - \omega_{-\mathbf{k}})^2 + (\gamma_0 + \gamma_\mathbf{k} + \gamma_{-\mathbf{k}})^2/\hbar^2} \ ,$$

$$G^{(-)}_{\mathbf{k},\mathbf{k}',\mathbf{k}+\mathbf{k}'} = \frac{-i(\omega_0 - \omega_\mathbf{k} - \omega_{\mathbf{k}'} + \omega_{\mathbf{k}+\mathbf{k}'}) + (\gamma_0 + \gamma_\mathbf{k} + \gamma_{\mathbf{k}'} + \gamma_{\mathbf{k}+\mathbf{k}'})/\hbar}{(\omega_0 - \omega_\mathbf{k} - \omega_{\mathbf{k}'} + \omega_{\mathbf{k}+\mathbf{k}'})^2 + (\gamma_0 + \gamma_\mathbf{k} + \gamma_{\mathbf{k}'} + \gamma_{\mathbf{k}+\mathbf{k}'})^2/\hbar^2} \ ,$$

$$G^{(+)}_{\mathbf{k},\mathbf{k}',\mathbf{k}+\mathbf{k}'} = \frac{i(\omega_0 - \omega_\mathbf{k} - \omega_{\mathbf{k}'} + \omega_{\mathbf{k}+\mathbf{k}'}) + (\gamma_0 + \gamma_\mathbf{k} + \gamma_{\mathbf{k}'} + \gamma_{\mathbf{k}+\mathbf{k}'})/\hbar}{(\omega_0 - \omega_\mathbf{k} - \omega_{\mathbf{k}'} + \omega_{\mathbf{k}+\mathbf{k}'})^2 + (\gamma_0 + \gamma_\mathbf{k} + \gamma_{\mathbf{k}'} + \gamma_{\mathbf{k}+\mathbf{k}'})^2/\hbar^2} \ .$$

Equation (A3) is now rearranged and becomes after some algebra

$$E(t) = -\left(\langle B^+B\rangle + \langle[B,B^+]\rangle\right)\left([A\rho_0(t),A^+] + [A,\rho_0(t)A^+] - i[A^+A,\rho_0(t)]\right)-$$
$$\langle B^+B\rangle\left([A^+\rho_0(t),A] + [A^+,\rho_0(t)A]\right).$$

In order to obtain the explicit form of the contributions to the master equation, we now replace $A$ and $B$ with their definition and calculate the expectation values of the "reservoir" variables: For $B = \sum_{\mathbf{k}\neq 0,} G^{(+)}_{\mathbf{k},-\mathbf{k}} W_{\mathbf{k},-\mathbf{k}}(P_\mathbf{k} P_{-\mathbf{k}})$ the expectation values of the k-modes lead to

$$\mathrm{Re}\langle[B,B^+]\rangle_1 + \mathrm{Re}\langle B^+B\rangle_1 = 2\sum_{\mathbf{k}\neq 0,}\mathrm{Re}\,G^{(+)}_{\mathbf{k},-\mathbf{k}}W_{\mathbf{k},-\mathbf{k}}^2\left(\langle P^+_\mathbf{k}P_\mathbf{k}\rangle+1\right)\left(\langle P^+_{-\mathbf{k}}P_{-\mathbf{k}}\rangle+1\right),$$

$$\mathrm{Im}\langle[B,B^+]\rangle_1 = 2\sum_{\mathbf{k}\neq 0,}\mathrm{Im}\,G^{(+)}_{\mathbf{k},-\mathbf{k}}W_{\mathbf{k},-\mathbf{k}}W_{\mathbf{k},-\mathbf{k}}\left(\langle P^+_\mathbf{k}P_\mathbf{k}\rangle+\langle P^+_{-\mathbf{k}}P_{-\mathbf{k}}\rangle\right),$$

$$\mathrm{Re}\langle B^+B\rangle_1 = 2\sum_{\mathbf{k}\neq 0,}\mathrm{Re}\,G^{(+)}_{\mathbf{k},-\mathbf{k}}W_{\mathbf{k},-\mathbf{k}}^2\langle P^+_\mathbf{k}P_\mathbf{k}\rangle\langle P^+_{-\mathbf{k}}P_{-\mathbf{k}}\rangle.$$

For $B = \sum_{\mathbf{k},\mathbf{k}',}G^{(+)}_{\mathbf{k},\mathbf{k}',\mathbf{k}+\mathbf{k}'}W_{\mathbf{k},\mathbf{k}'}P^+_{\mathbf{k}+\mathbf{k}'}P_\mathbf{k}P_{\mathbf{k}'}$ we obtain

$$\mathrm{Re}\langle[B,B^+]\rangle_2 + \mathrm{Re}\langle B^+B\rangle_2 = \sum_{\mathbf{k},\mathbf{k}',}\mathrm{Re}\,G^{(+)}_{\mathbf{k},\mathbf{k}',\mathbf{k}+\mathbf{k}'}W_{\mathbf{k},\mathbf{k}'}\left(W_{\mathbf{k},\mathbf{k}'}+W_{\mathbf{k}',\mathbf{k}}\right)\times$$

$$\left[\langle P^+_{\mathbf{k}+\mathbf{k}'}P_{\mathbf{k}+\mathbf{k}'}\rangle\left(\langle P^+_\mathbf{k}P_\mathbf{k}\rangle+1\right)\left(\langle P^+_{\mathbf{k}'}P_{\mathbf{k}'}\rangle+1\right) - \left(\langle P^+_{\mathbf{k}+\mathbf{k}'}P_{\mathbf{k}+\mathbf{k}'}\rangle+1\right)\langle P^+_\mathbf{k}P_\mathbf{k}\rangle\langle P^+_{\mathbf{k}'}P_{\mathbf{k}'}\rangle\right],$$

$$\mathrm{Im}\langle[B,B^+]\rangle_2 = \sum_{\mathbf{k},\mathbf{k}',}\mathrm{Im}\,G^{(+)}_{\mathbf{k},\mathbf{k}',\mathbf{k}+\mathbf{k}'}W_{\mathbf{k},\mathbf{k}'}\left(W_{\mathbf{k},\mathbf{k}'}+W_{\mathbf{k}',\mathbf{k}}\right)\times$$

$$\left[\langle P^+_{\mathbf{k}+\mathbf{k}'}P_{\mathbf{k}+\mathbf{k}'}\rangle\left(\langle P^+_\mathbf{k}P_\mathbf{k}\rangle+\langle P^+_{\mathbf{k}'}P_{\mathbf{k}'}\rangle+1\right) - \langle P^+_\mathbf{k}P_\mathbf{k}\rangle\langle P^+_{\mathbf{k}'}P_{\mathbf{k}'}\rangle\right],$$

$$\mathrm{Re}\langle B^+B\rangle_2 = \sum_{\mathbf{k},\mathbf{k}',}\mathrm{Re}\,G^{(+)}_{\mathbf{k},\mathbf{k}',\mathbf{k}+\mathbf{k}'}W_{\mathbf{k},\mathbf{k}'}\left(W_{\mathbf{k},\mathbf{k}'}+W_{\mathbf{k}',\mathbf{k}}\right)\times$$

$$\left[\langle P^+_{\mathbf{k}+\mathbf{k}'}P_{\mathbf{k}+\mathbf{k}'}\rangle\langle P^+_\mathbf{k}P_\mathbf{k}\rangle\langle P^+_{\mathbf{k}'}P_{\mathbf{k}'}\rangle + \langle P^+_\mathbf{k}P_\mathbf{k}\rangle\langle P^+_{\mathbf{k}'}P_{\mathbf{k}'}\rangle\right].$$

The terms containing the imaginary parts of the expectation values $\langle B^+B\rangle$ represent energy shifts due to the interaction with the reservoir that in the following are absorbed into the shifted polariton frequency $\hat{\omega}_0$ and self –interaction coefficient

FIGURE CAPTIONS

Fig. 1

A sketch of the non-resonant excitation, the exciton formation and relaxation to $k=0$ in a QW embedded in a microcavity.

Fig. 2
Gain and amplification of the mode $k=0$ as a function of the polariton density in the bottleneck i. e. of the strength of the external pump for a CdTe QW.

Fig. 3
Plot of the polariton distribution $\rho(n)$ for a CdTe QW for different polariton densities in the bottleneck.

Fig.4
Plot of first moment of the polariton distribution $\rho(n)$ for a CdTe QW as a function of the exciton density in the bottleneck. It clearly shows amplification at threshold. For comparison the same first moment is presented for a one-mode laser.

Fig. 5
Plot of the second normalized moment $M_2$ of the polariton distribution $\rho(n)$ for a CdTe QW as a function of the exciton density in the bottleneck. For comparison the same second normalized moment is presented for a one-mode laser.

Fig. 6
Plot of the linewidth of the polariton emission for a CdTe QW as a function of the exciton density in the bottleneck. The linewidth decreases and attains its minimum at threshold.

Fig. 7
Plot of the polariton amplitude for a CdTe QW as a function of the exciton density in the bottleneck. For comparison the mean field approximation is also indicated

FIGURES

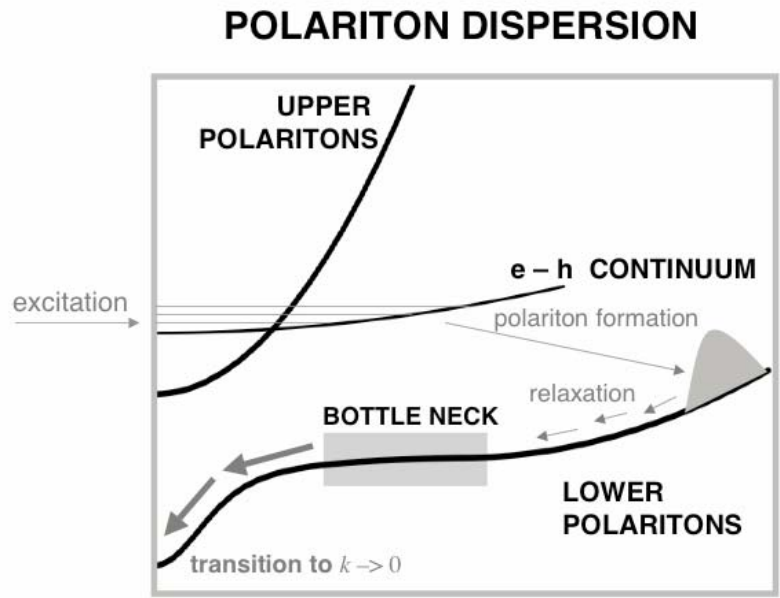

Fig. 1

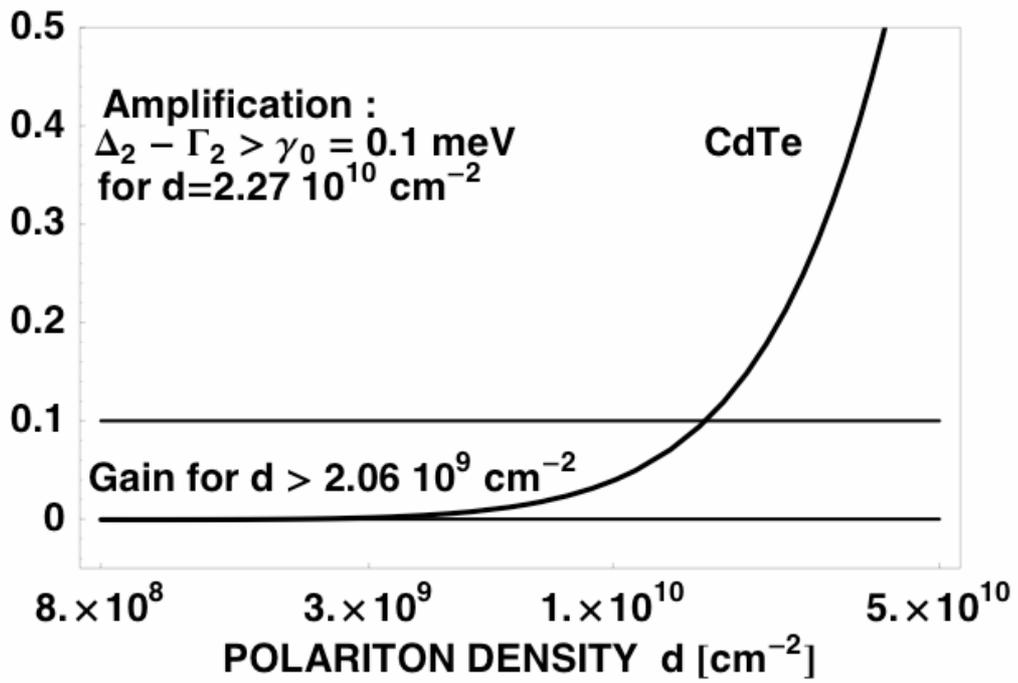

Fig. 2

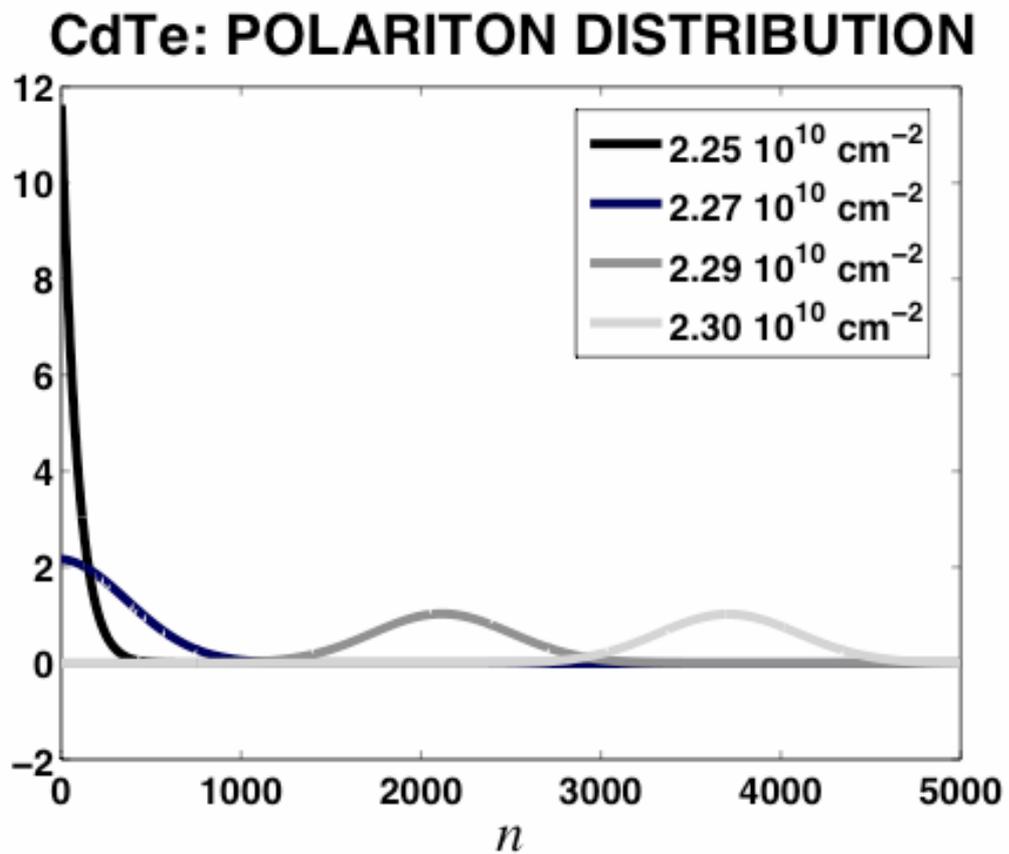

Fig. 3

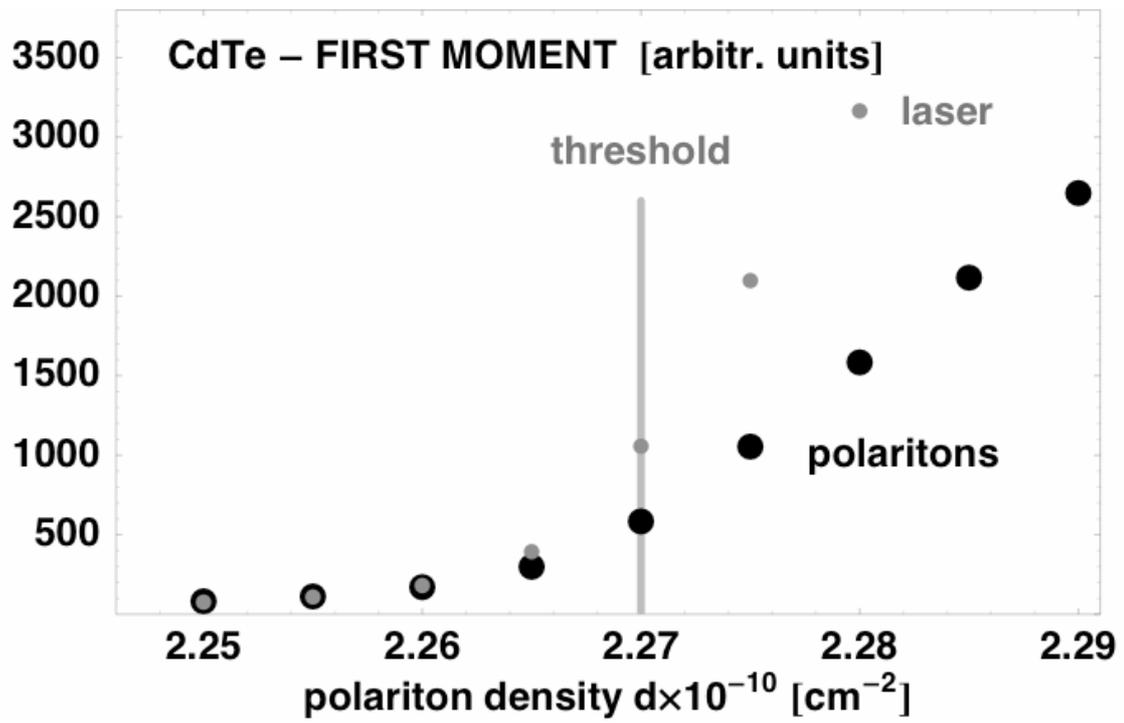

Fig. 4

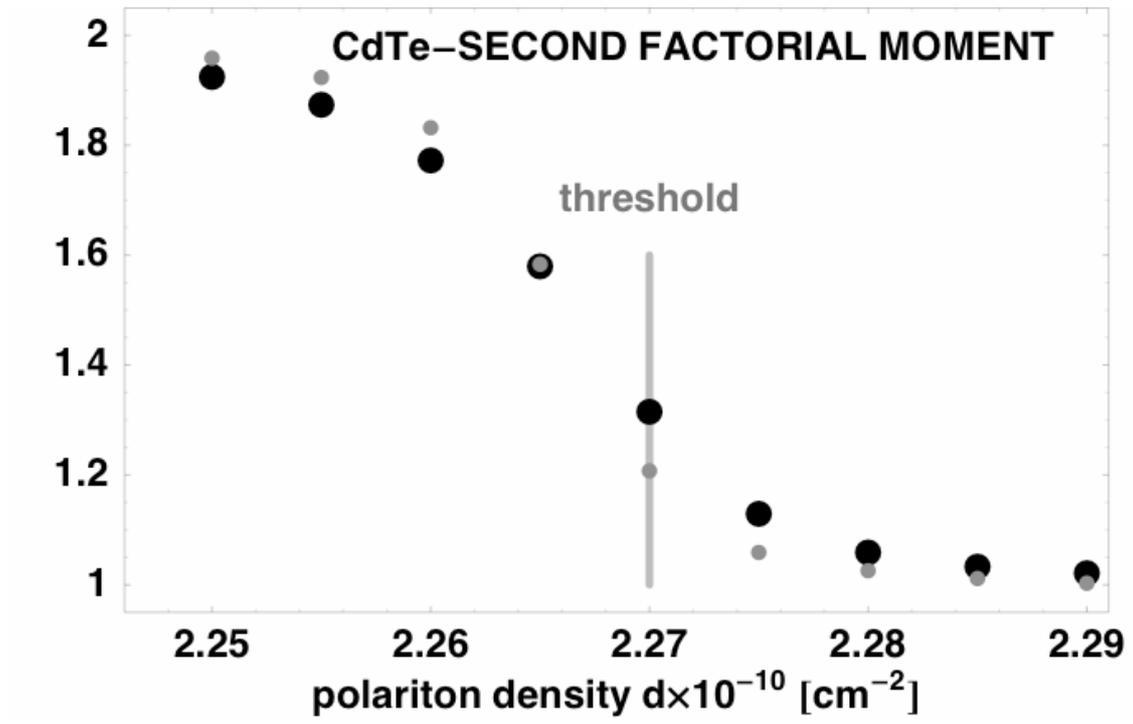

Fig. 5

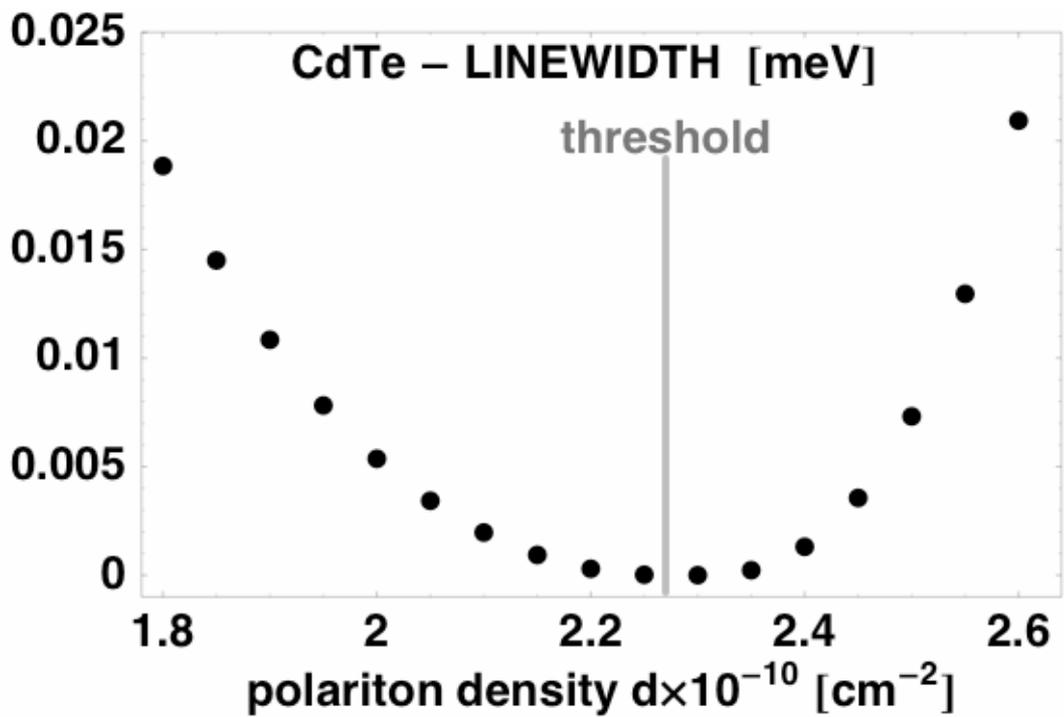

Fig. 6

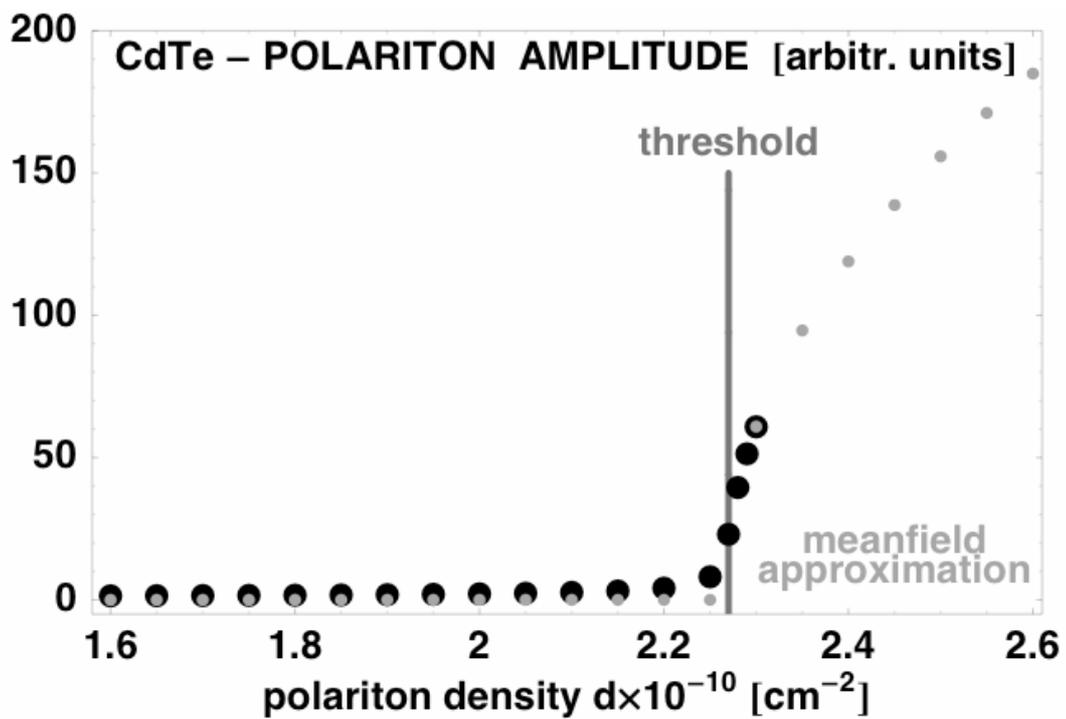

Fig. 7